**ORIGINAL PAPER**

# The design and performance of GRD onboard the GECAM satellite

Z. H. An[1] · X. L. Sun[1,2] · D. L. Zhang[1] · S. Yang[1] · X. Q. Li[1] · X. Y. Wen[1] · K. Gong[1] · X. H. Liang[1] · X. J. Liu[1] ·
Y. Q. Liu[1] · Y. G. Li[1] · S. L. Xiong[1] · Y. B. Xu[1] · Fan Zhang[1] · X. Y. Zhao[1] · C. Cai[1] · Z. Chang[1] · G. Chen[1] · C. Chen[1] ·
Y. Y. Du[1] · P. Y. Feng[1] · M. Gao[1] · R. Gao[1] · D. Y. Guo[1] · J. J. He[1] · D. J. Hou[1] · C. Y. Li[1] · G. Li[1] · L. Li[1] · X. F. Li[1] ·
M. S. Li[1] · F. J. Lu[1] · H. Lu[1] · B. Meng[1] · W. X. Peng[1] · F. Shi[1] · H. Wang[1] · J. Z. Wang[1] · Y. S. Wang[1] · H. Z. Wang[1] ·
X. Wen[1] · S. Xiao[1] · Y. P. Xu[1] · J. W. Yang[1] · Q. B. Yi[1,3] · S. N. Zhang[1] · C. Y. Zhang[1] · C. M. Zhang[1] · Fei Zhang[1] ·
Y. Zhao[1,4] · X. Zhou[1]

**Abstract**

**Background**  Each GECAM satellite payload contains 25 gamma-ray detectors (GRDs), which can detect gamma-rays and particles and can roughly localize the Gamma-Ray Bursts (GRBs). GRD was designed using lanthanum bromide (LaBr$_3$) crystal as the sensitive material with the rear end coupled with silicon photomultiplier (SiPM) array for readout.

**Purpose**  In aerospace engineering design of GRD, there are many key points to be studied. In this paper, we present the specific design scheme of GRD, the assembly and the performance test results of detectors.

**Methods**  Based on Monte Carlo simulation and experimental test results, the specific schematic design and assembling process of GRD were optimized. After being fully assembled, the GRDs were conducted performance tests by using radioactive source and also conducted random vibration tests.

**Result and conclusion**  The test results show that all satellite-borne GRDs have energy resolution <16% at 59.5 keV, meeting requirements of satellite in scientific performance. The random vibration test shows that GRD can maintain in a stable performance, which meets the requirement of spatial application.

## Introduction

Gravitational wave high-energy Electromagnetic Counterpart All-sky Monitor (GECAM) is a small space science detection project specifically designed to detect gravitational wave high-energy electromagnetic counterpart (referred to as gravitational wave gamma-ray burst) for the purpose of making research on gravitational wave astronomy [1]. Gravitational wave high-energy electromagnetic counterpart refers

✉ Z. H. An
  anzh@ihep.ac.cn

✉ X. Q. Li
  lixq@ihep.ac.cn

1  CAS, Institute of High Energy Physics, Beijing 100049, China

2  State Key Laboratory of Particle Detection and Electronics, Beijing 100049, China

3  Xiangtan University, Xiangtan 411105, China

4  Beijing Normal University, Beijing 110875, China

to the celestial source of electromagnetic radiation homologous to a gravitational wave in time and space, that is, the light emitted by the gravitational wave source. According to existing researches, the binary mergers involving neutron stars (e.g., neutron star-neutron star or neutron star-black hole mergers) not only produces gravitational waves, but also is often accompanied by radiation in the bands of X/γ rays, soft X-rays, optics, and radio. Among them, X/γ rays often appear in form of gamma-ray burst [2]. In aspect of the detection of high-energy electromagnetic counterpart, almost all running high-energy detectors including Fermi/GBM [3], Swift/BAT [4], INTEGRAL/SPI-ACS [5], Konus-Wind [6], and CALET/CGBM [7], both in China and foreign countries, have invested large amount of resources to search and research. Those satellites and the gamma-ray detectors on GECAM are used for observing the high-energy electromagnetic counterpart in this research.

The GECAM project consists of two small satellites running in the same orbital plane but opposite orbital phases

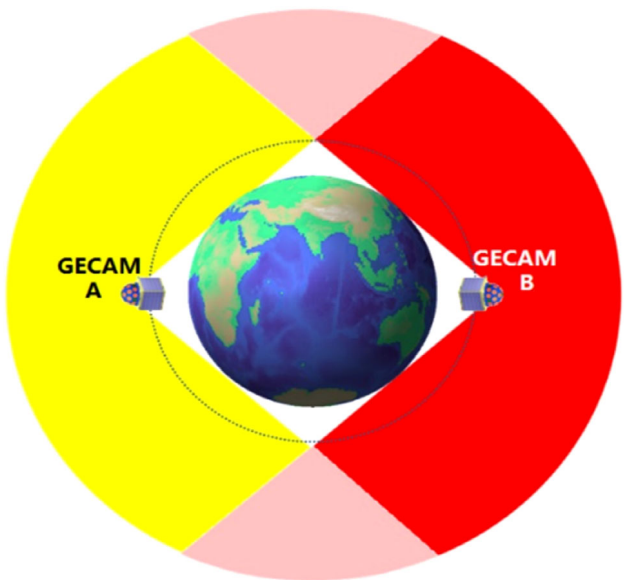

**Fig. 1** Schematic diagram of field of satellites

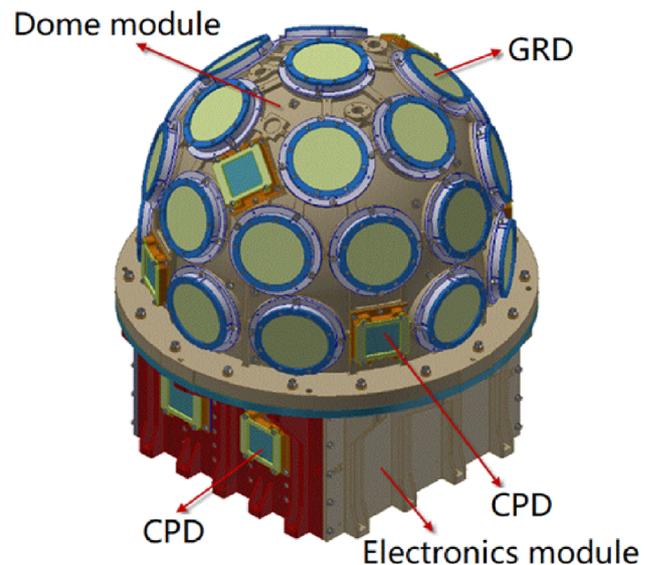

**Fig. 2** Structure of payloads of GECAM

(Fig. 1) to make all-sky complete coverage monitoring [8]. The payloads of each satellite include 25 gamma-ray detectors (GRDs), 8 charged particle detectors (CPDs), and Electronics module (EBOX) (Fig. 2, where the round one represents GRD and the square one represents CPD). The 25 GRD modules are all installed in the dome of the satellite, pointing to different directions. Such a design provides GRD with wide monitoring field of view. Among the eight CPDs, six CPDs are installed in the dome of the satellite and two remained CPDs are installed on the side of payload Electronics module. CPD is used to measure the flux changes of charged particles in the space environment, and can also be used to identify gamma-ray burst and charged particles event in the space in combination with GRD. Positioned in payload Electronics module, EBOX is used for acquiring and processing GRD and CPD data, realizing on-orbit triggering and positioning of gamma-ray burst, completing data storage, packaging, transmission and the interaction with satellite platform, and providing power supply for the entire payload.

As the main detector of GECAM satellite payload with large area and field of view and high detection efficiency, GRD can be used for triggering and positioning of gamma-ray burst and observing the light-curve and energy spectrums. Its main functions are as follows:

1. To conduct all-sky monitoring on gamma-ray burst within the wave band from X-ray to gamma ray (covers the energy range from 8 to 2000 keV);
2. To quickly trigger and position gamma-ray burst in the field of view;

3. To measure the light-curve and energy spectrums of various gamma-ray bursts;
4. To detect gravitational wave gamma-ray burst in combination with ground gravitational wave detectors;
5. To identify gamma-ray burst and charged particle burst together with CPD payload.

According to the scientific needs of satellite, the detection energy range of GRD was selected as 8 keV–2 MeV to facilitate measuring wide energy spectrum of gamma-ray burst and even obtaining enough high gamma photon statistical magnitude at the low energy end, so as to further reduce the statistical error and improve the positioning accuracy. In order to meet the scientific needs of GECAM and the requirements of small satellite platforms, GRD should be small and low-power while taking low-energy gamma-ray detection into full consideration. To meet this requirement in the design of GRD, a method of combined use of $LaBr_3$ crystal and silicon photomultiplier (SiPM) was designed for readout since $LaBr_3$ crystal had high density and light yield, and short luminescence decay time and SiPM instead of the traditional Photomultiplier Tube (PMT) had simple and compact structure and low power consumption and was easy to be miniaturized.

In the early stage of the project development, the basic performance of $LaBr_3$ crystal coupled SiPM had been tested [9, 10]. The test result shows that this design can meet the index requirement of satellite. However, there are some detailed problems. For instance, $LaBr_3$ crystal should be fully encapsulated in an aluminum casing due to the hygroscopy of the material; meanwhile, the coupling between $LaBr_3$ crystal and SiPM should be optimized; and the gain of SiPM varies with the change in temperature, which should be corrected.

**Table 1** Main performance indexes of GRD

| Parameters | Value |
|---|---|
| Detection energy range | 8 keV–2 MeV |
| Detection area | >40 cm² (for each GRD) |
| Dead-time | <4 μs (normal event) ~ 70 μs (overflow event) |
| Energy resolution | <18%@59.5 keV |
| Detection efficiency for Gamma-rays | >60%@8 keV |

**Table 2** Main scintillation properties for widely used scintillator media

| Crystal | LaBr₃(ce) | NaI(Tl) | CsI(Tl) | BGO |
|---|---|---|---|---|
| Density (g/cm³) | 5.08 | 3.67 | 4.51 | 7.13 |
| Melting Point (°C) | 783 | 651 | 621 | 1050 |
| Decay Time (ns) | 16 | 230 | 1250 | 300 |
| Light Yield (Ph/MeV) | 78,000 | 38,000 | 65,000 | 8500 |
| Luminescence (nm) (at Peak) | 358 | 415 | 550 | 480 |
| Hygroscopic | yes | yes | slightly | no |

## The encapsulated LaBr₃ crystal unit

LaBr₃ crystal is one of the most excellent mass-produced scintillation crystals at present (Table 2) [11–13]. Among the crystals that are widely used, LaBr₃ at 662 keV has 1.65 times the light yield of NaI(Tl) crystal and also stands out by a more than an order of magnitude faster decay time compared to NaI(Tl). The peak wavelength emitted LaBr₃ it is 380 nm, which well matches with the sensitive wavelength of SiPM (with photon detection efficiency ranging within 30–40%). Basically, the largest size that can be achieved in international mass production of LaBr₃ crystal is 3 inches. Combined with scientific goals, the size of LaBr₃ crystal for the said GRD was finalized at 76 mm in diameter and 15 mm in thickness. Because the radioactive isotope ²²⁷Ac and its daughters and the radioactive isotope ¹³⁸La in the crystal have radioactive background, the crystal has two background characteristic peaks, respectively, at 37.4 keV and 1470 keV [14], which can be used for on-orbit energy calibration [15].

Being highly hygroscopic, LaBr₃ crystal can chemically react with water in the air to generate lightless bromic acid and other substances. Thereby, once deliquescence occurs, a dead layer will appear on the surface and directly affect the detection of low-energy X-rays; meanwhile, the light output of the crystal will also deteriorate. Hence, LaBr₃ crystal should be fully encapsulated. The appearance of the unit is as shown in Fig. 4. The assembly of LaBr₃ crystal packaging unit was manufactured by Beijing Glass Research Institute, China. The components from top to bottom are, respectively, beryllium window, aluminum shell, LaBr₃ crystal, and quartz window bonded together by using sealant so as to isolate the crystal from the atmospheric water vapor and maintain it dry.

Considering the packaging technique, since LaBr₃ crystal is hygroscopic and needs to withstand the test in aerospace environment, the crystal needs to be packaged and wrapped with a reflective layer; and shock absorption measures should be taken between the crystal and the aluminum shell, so as to prevent the crystal from splintering due to the vibration during rocket launch. For detecting low-energy X-ray (<10 keV), the incident window of the crystal housing unit should be as thin as possible. Provided that the structural strength is guaranteed, beryllium window is the best choice.

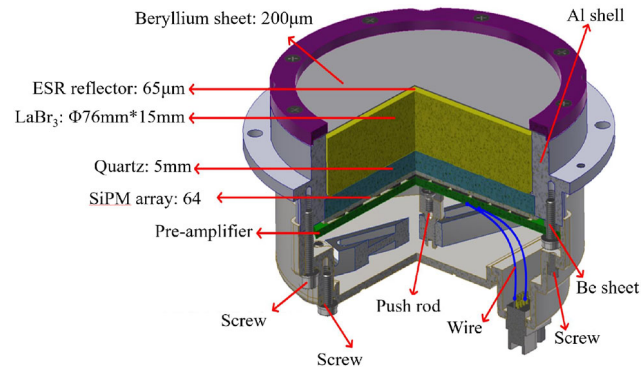

**Fig. 3** GRD module structure

Labels on figure: Beryllium sheet: 200μm; Al shell; ESR reflector: 65μm; LaBr₃: Φ76mm*15mm; Quartz: 5mm; SiPM array: 64; Pre-amplifier; Screw; Push rod; Screw; Wire; Be sheet; Screw

Those problems should be provided with specific solutions in the project development. In aerospace engineering design and specific circuit design of GRD, there are still many key points to be studied. This paper describes the specific design scheme of GRD onboard the GECAM satellite, the assembly of detector and the results of its performance test on radioactive source. According to the final radioactive source test and random vibration test results, GRD satisfies the conditions for being loaded on the satellite due to its stable quality and excellent performance.

## The design of GRD

According to the task demands of GECAM, requirements for the main performance indexes of GRD are as shown in Table 1. GRD adopted LaBr₃ (3 in. in diameter and 15 mm thickness) and the silicon photomultiplier tube (SiPM) array. The GRD onboard the GECAM satellite is mainly composed of the encapsulated LaBr3 crystal unit, SiPM array, pre-amplification electric circuit electronics system and mechanical structure, its internal structure is shown in Fig. 3.

**Fig. 4** LaBr$_3$ crystal housed in an aluminum hermetically sealed container with a quartz optical window on one end

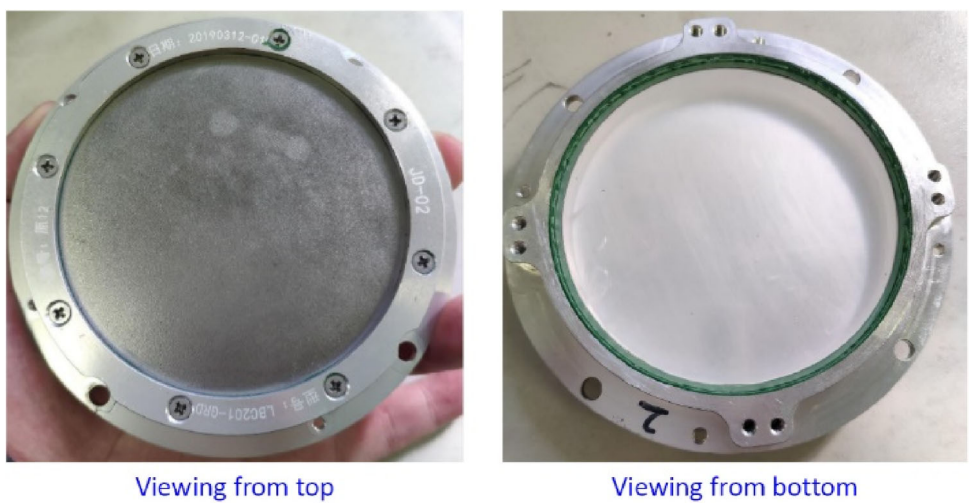

Viewing from top      Viewing from bottom

For beryllium window with thickness of 200 µm, the penetration rate of X-ray at 5 keV is greater than 60%, which meets the low-energy X-ray detecting requirement of GRD. Meanwhile, adopting beryllium window also enhances the difficulty in packaging of the crystal. The light output surface of the crystal should be coupled with quartz glass so as to both getting good sealing effect and getting the scintillation light generated by the crystal transmitted out through the quartz window, for main reason that quartz glass has good penetration rate within the wavelength range of 320–450 nm.

The key to sealing and long-term stable performance of the LaBr$_3$ crystal lies in a perfect crystal packaging technique, which is a decisive factor to prevent the crystal from deliquescing. The most important research content is to verify the reliability of the technique. Over a series of tests in mechanical, thermal, vacuum, and other astronautic conditions and by improving the technique again and again, the LaBr$_3$ crystal packaging unit finally passed all the tests.

## SiPM array of GRD

Silicon photomultiplier (SiPM) consists of multiple Avalanche Photo-Diode (APD) arrays working in Geiger mode. Each APD is a pixel and will output a charge pulse signal after receiving a photon [16]. The sum of the charges output from all pixels is directly proportional to the total number of photons detected by SiPM. SiPM has superior photon counting ability and weak light signal detection ability, insensitivity to magnetic fields, low bias voltage, high gain, small size, low power consumption, high quantum efficiency, and simple structure, and is convenient for being integrated with detector.

The SiPM for GRD adopted the MicroFJ-60035-TVS produced by ON Semiconductor (formally SensL) [17]. Each small chip of SiPM was sized 6.07 × 6.07 mm and contained 22,292 microcells (each microcell was sized 35 µm). For 3-inch LaBr$_3$ crystal, each crystal was read out by using 64 SiPM chips. To ensure collecting uniform flare light, special design of SiPM array of GRD was made on the arrangement of the SiPM chips (Fig. 5). In such design, good light collection efficiency can be got by using SiPM chips as few as possible. In contrast, using over many SiPM chips may increase the electric current and power consumption of the detector and also add unnecessary noise. The 64 SiPM chips were divided into two groups to ensure getting reliable power supply. The detailed design is introduced in other papers about the project.

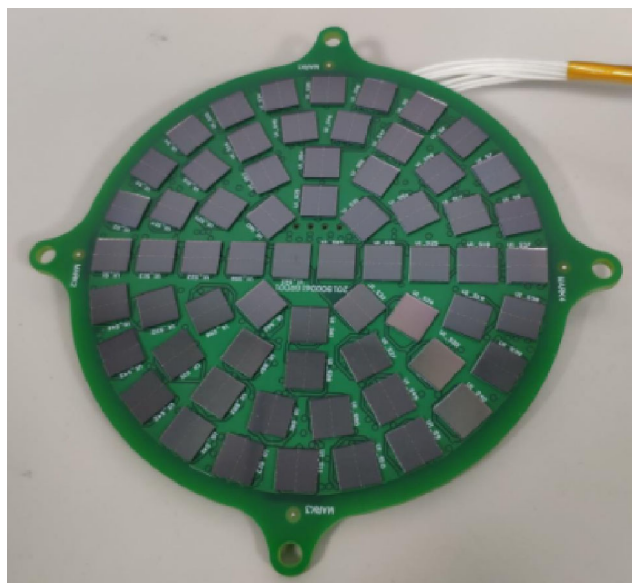

**Fig. 5** SiPM array of GRD

As SiPM is a semiconductor device, its noise level (thermal noise counting rate) increases significantly as the temperature rises. Therefore, the lower the operating temperature is, the more it is conducive to suppress noise. In view of this, the SiPM was made thermal design, with the operating temperature designated at $-20\pm3$ °C, which could greatly reduce the noise of SiPM.

**Fig. 6** Gain temperature control schematic of SiPM

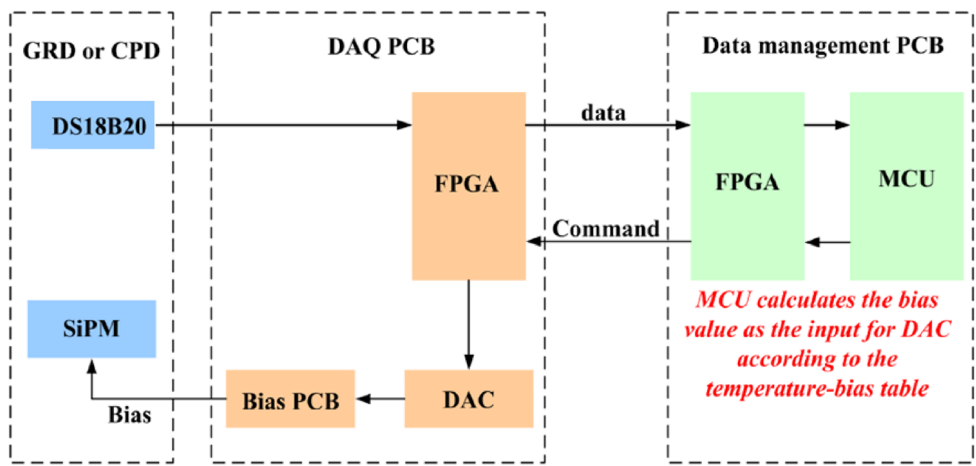

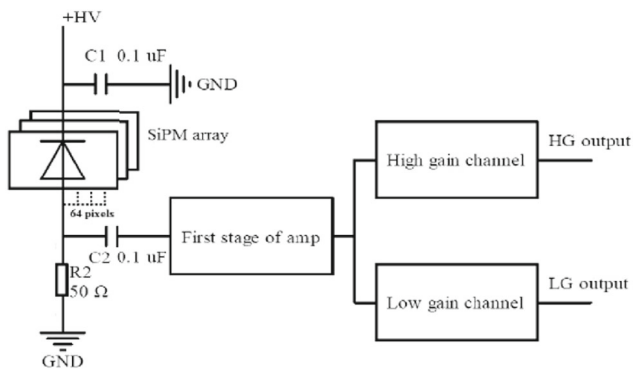

**Fig. 7** Block scheme of SiPM array and preamplifier [9]

Located on the surface of satellite, GRD will surely have temperature fluctuations when working on orbit. Since the gain coefficient of SiPM is about 3%/ °C and sensitive to temperature, temperature change will affect the gain stability of GRD, resulting in large error in the measured energy value of gamma ray. To solve the problem of temperature coefficient of SiPM, a temperature compensation circuit was added in the circuit design. After entering the orbit, satellite acquired the temperature of SiPM in time by using temperature sensor. Then payload data acquisition and control circuit timely adjusted the bias voltage of SiPM as per the temperature to modify the gain temperature of SiPM and maintain the gain of SiPM stable. The detailed principles are shown in Fig. 6. The temperature data from the Digital temperature sensor (DS18B20) will be transmitted to the FPGA, where it will be again transmitted to the Data management PCB through internal LVDS interface. The MCU will calculate the bias value as the input for the DAC according to the temperature-bias table of SiPM.

## Pre-amplification electric circuit

Because GRD needed to measure the energy within a large dynamic range from keV to MeV, the pre-amplification electronic reader of SiPM adopted high- and low-gain readout methods to process SiPM signals and read out large dynamic range of signals in the complete energy range. The SiPM gain of GRD was about $10^6$. In the design of preamplifier, the signal output by the SiPM array needed to pass through the DC blocking capacitor, the first-order preamplifier, and the second-order preamplifier with high and low gains successively and finally be output to data acquisition circuit and made Analog-to-Digital (A/D) conversion. Block diagram of the pre-amplification electric circuit of GRD is as illustrated in Fig. 7.

The SiPM output signal was amplified by two orders of operational amplifiers. Each order adopted the same operational amplification principle but different feedback resistances. The first-order amplification times of amplifiers were both 1. The second-order amplification times were different: with high gain, the signal was amplified by 8.3 times the gain and converted into differential signal; while with low gain, the signal was amplified by once the gain and then directly output. The signal output from preamplifier was transmitted to data acquisition circuit and then was molded, amplified, and subjected to A/D conversion. Finally, the energy range covered in high-gain channel of GRD was 3–300 keV and that in low-gain channel was 40 keV–4 MeV.

In terms of saturation signal processing, the GRD was designed with electronic components having circuit protection functions. Additionally, trigger logic was set on the data acquisition board to prevent saturation signal from being acquired for many times due to long duration. Hence, the influence brought about on the detection by saturation signal was reduced. Figure 8 shows the intrinsic activity energy spectra of LaBr$_3$ as acquired by DAQ. The characteristic

**Fig. 8** Tested intrinsic activity spectra of the GRD module

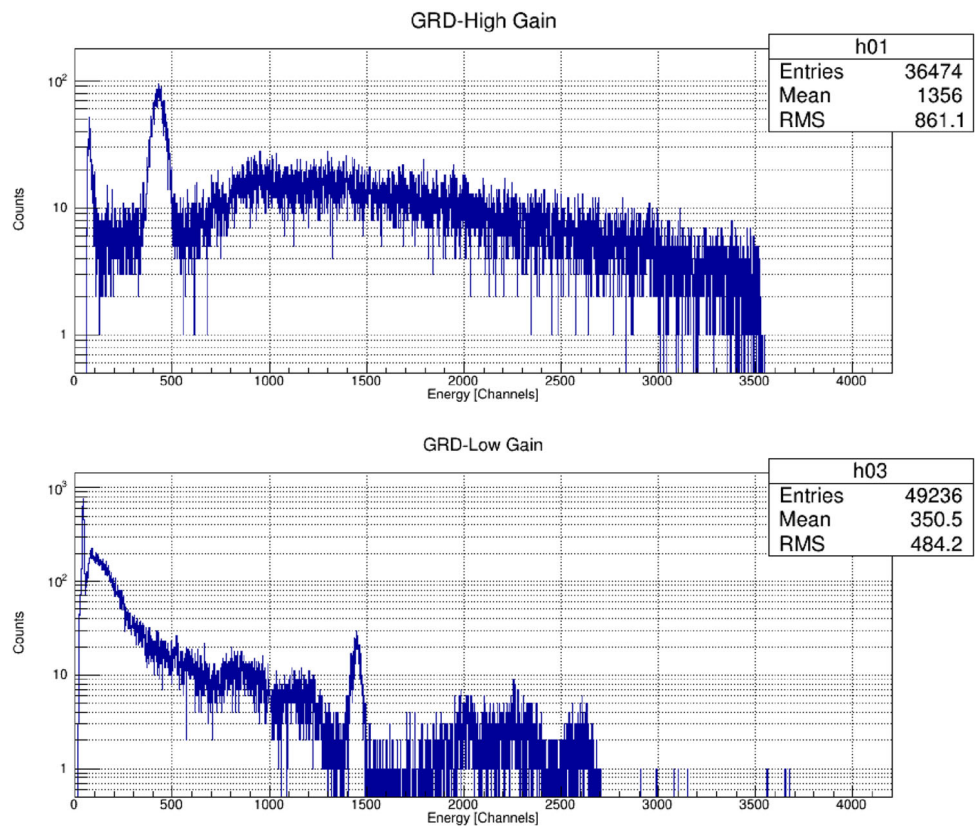

peaks of LaBr$_3$ crystal at 37.4 keV and 1470 keV are clearly visible.

## Assembly of GRD

The schematic diagram of GRD module structure is presented in Fig. 9. Pasting Tyvek reflective film in the gap between SiPMs can effectively increase the collection of scintillator light and improve the energy resolution of GRD. LaBr$_3$ crystal housing unit was coupled with SiPM by using 1 mm-thick optical silicone pad, which also contributed to buffering and shock absorption in addition to increasing light acquisition efficiency. As the size of pre-amplification electric circuit was large, a push rod was added in the center of the circuit board to ensure that the SiPM was close to the silicone pad. After getting the circuit board installed, the rear end transferred certain torsion to the push rod so as to get the SiPM between circuit boards well fitted to the silicone pad and enhance the light acquisition efficiency. The design in mechanical structure also took the requirements for light-free and electromagnetic compatibility into account. After being assembled, the detector was exhausted to vacuum state so as to get rid of air bubbles between the silicone pad and SiPM and improve the efficiency of scintillator light collection.

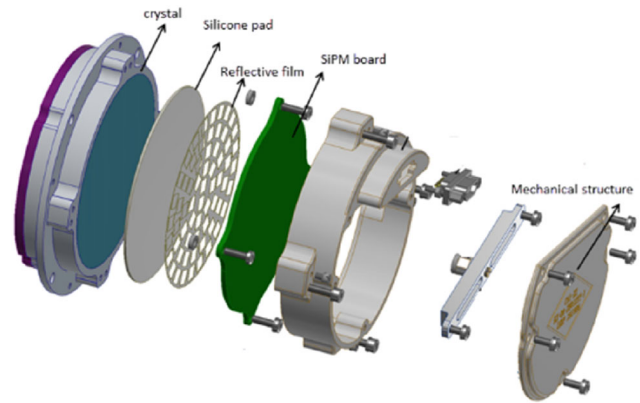

**Fig. 9** Schematic diagram of GRD

## Simulation and experimental testing of GRD

In this section, a simulation analysis was made on the full-scale model of GRD. And whether the GRD satisfied with the scientific and engineering target requirements of GECAM satellites or not was studied in combination with radioactive source test and random vibration test on the detector.

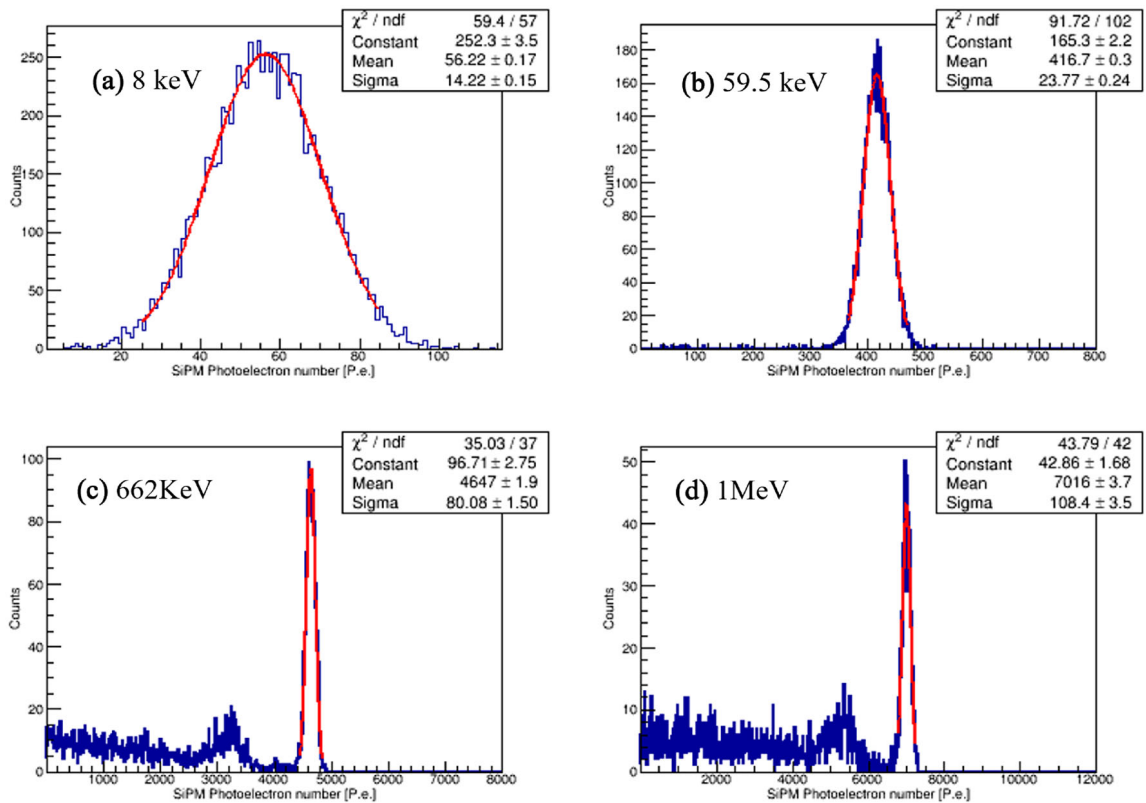

**Fig. 10** The detector energy spectrums of GRD at simulated gamma ray incident energies of 8 keV (**a**), 59.5 keV (**b**), 662 keV (**c**), and 1 MeV (**d**), respectively

**Fig. 11** The detection efficiency curve of GRD Full-energy peak obtained by simulation

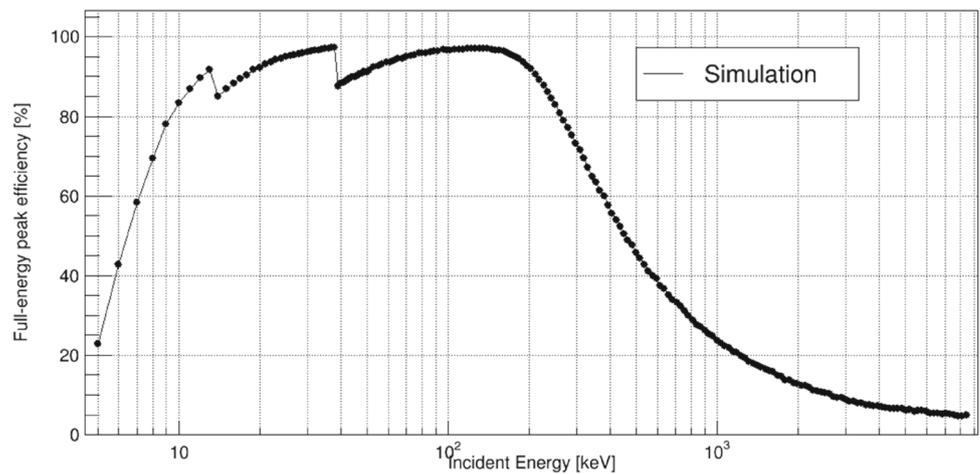

**Fig. 12** GRD radioactive source test schematic diagram

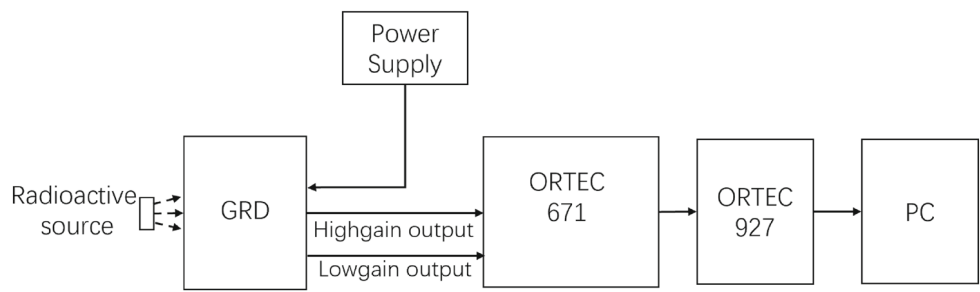

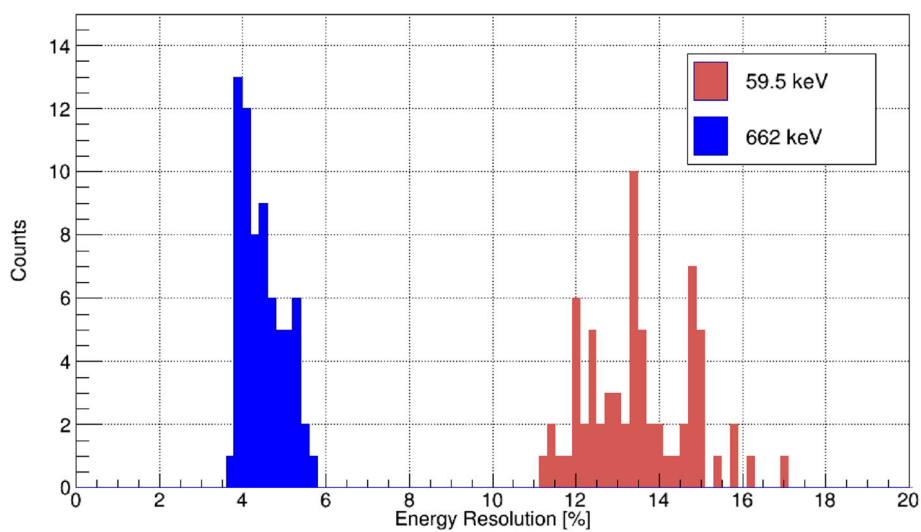

**Fig. 13** The radioactive source test results of energy resolution of GRDs

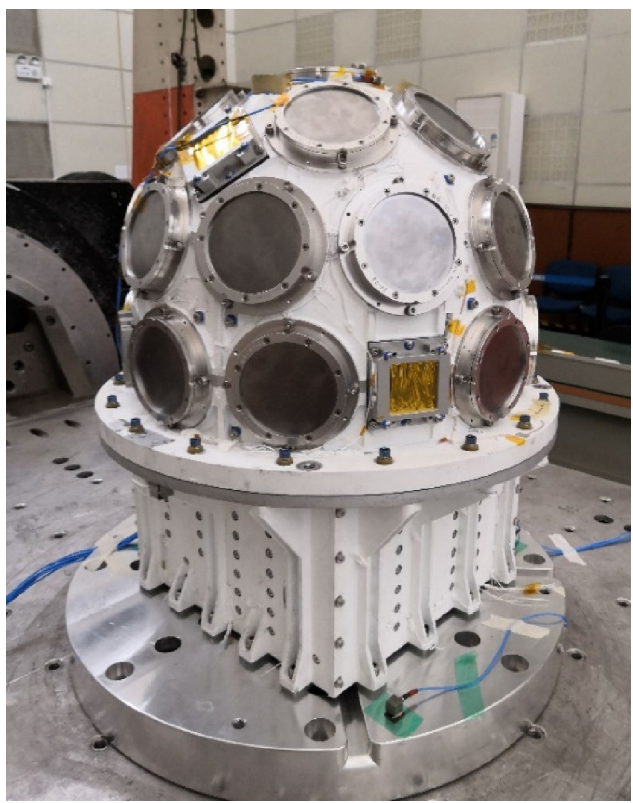

**Fig. 14** Photo of GRD Random vibration test

## Simulation analysis

The scientific requirement for the lower limit of GRD energy is 8 keV, and the expansion requirement is 6 keV. The beryllium window used should be only 200 μm in thickness in order to allow X-ray of a few keV to pass through effectively. According to the given physical process, geant4 was used to count the deposited energy of the particles on the crystal after entering the detector. Moreover, a simulation analysis was made on the scintillator light transmission, providing the energy spectrum and energy resolution of gamma ray at different energies.

Energy resolution is an important performance index of GRD. Figure 10 illustrates the energy spectrums of GRD when the energies of simulated gamma rays (vertical incidence) are 8 keV, 59.5 keV, 662 keV, and 1 MeV, respectively. As revealed in the simulation results, the number of photoelectrons received by SiPMs increase linearly with the increase in incident energy of the particles. When the ray has low energy, the number of photoelectrons received by SiPMs are less so that the energy resolution is not good; with the increase in energy, the energy resolution of GRD gets better. When the energy is 59.5 keV, the energy resolution result obtained by simulation can reach 13.4%, which is better than the index requirement when the energy resolution is 18% at 59.5 keV.

Figure 11 presents the detector efficiency curve of GRD full-energy peak obtained by simulation. As shown, the results imply that the full-energy peak detection efficiency of GRD is about 70% at 8 keV, which meets the index requirement for being larger than 50% at 8 keV. Affected by beryllium window and other incident direction material, low-energy gamma ray has slightly low detection efficiency. However, with the increase in energy, the efficiency also increases. After the energy reaches more than 126 keV, the full-energy peak detection efficiency of GRD reduces gradually with the increase in penetrability of gamma ray. When the energy reaches 2 MeV, the full-energy peak detection efficiency of GRD is about 12%.

## Radioactive source test

Because LaBr3 crystals are produced in three batches and there are some differences in SiPM array and pre-amplification electronic, the GRDs performance is different.

**Table 3** Random vibration test Parameter

| Frequency Range(Hz) | Input PSD(g²/Hz) Verification Level |
|---|---|
| 20–100 | + 3 dB/oct |
| 100–500 | 0.08 |
| 500–1000 | 0.06 |
| 1000–2000 | − 12 dB/oct |
| Total Acceleration (RMS) | 10.5 g |
| Experiment Duration | 1 min |
| Directions | X, Y, Z |

At normal temperature and pressure, the performances of all the 68 GRDs assembled were tested on $^{241}$Am and $^{137}$Cs radioactive sources. The schematic diagram of GRD radioactive source is as shown in Fig. 12. The signal output by the detector was amplified by the main amplifier plug-in ORTEC 671, with gain of 14 and molding time of 0.5 μs, and the Gaussian molding function was enabled. The impulse amplitude spectrum output by the main amplifier plug-in was collected by multi-channel analyzer ORTEC ASPEC-927. The test result is as presented in Fig. 13.

According to the test results, the energy resolution of GRD to $^{241}$Am radioactive source is generally better than 16%, which satisfies the design requirement of satellite. Finally, 50 GRDs were selected by energy resolution from the 68 GRDs assembled and installed on two satellites of GECAM.

### Random vibration test

For the sake of verifying the reliability and structural stability properties of GRD, a verification level random vibration test was made on the 50 GRDs to be installed on satellite (Fig. 14) under the same mechanical test conditions as formal prototype of payload. The GRDs were installed in applicable positions on the dome and then subjected to random vibration test under mechanical conditions of the entire payload. The test conditions are demonstrated in Table 3.

During the test, only appearance of the detector was observed and no power was supplied; the performance of GRD was tested on radioactive source before and after the test, respectively, and then a comparison was made on the results.

As indicated in the test results (Fig. 15), before and after the test, the GRD maintains in stable performance, with relative change in peak less than 3% and that in energy resolution even smaller, which meets the requirements for use in aerospace.

### Conclusion

Based on Monte Carlo simulation and experimental testing, the specific design of GRD onboard the GECAM satellite was determined and the detailed composition and assembly of GRD were provided. After getting all GRDs assembled, performance of the detectors was tested on radioactive source and a random vibration test was made on the detectors to be installed on satellite. The results indicate that all GRDs installed on satellite have energy resolution < 16% at 59.5 keV, which meets the performance design requirement. Before and after the random vibration test, the relative change in peak of radioactive source of GRD is less than 3%, presenting a stable performance. Further, it is also needed to conduct performance calibration test on GRDs on the ground and scale the E–C relation, energy resolution, and other performances of the GRDs to prepare for the inversion of on-orbit

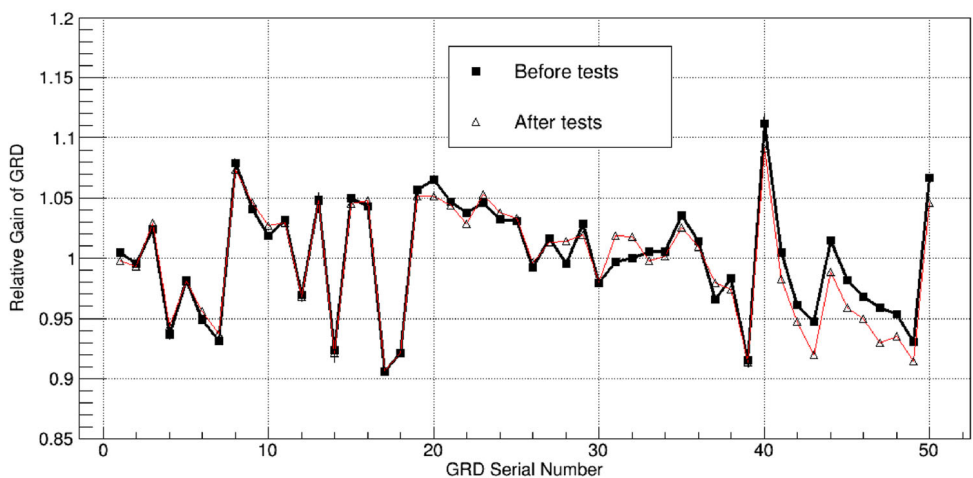

**Fig. 15** Before and after GRD Random vibration test, the change of 59.5 keV Full-energy Peak

data. This part of the work will be discussed in subsequent papers.

**Acknowledgments** The authors wish to thank Researcher Zhang Chunsheng, and thank Beijing Glass Research Institute for production and supplying of sealed LaBr$_3$ crystals for GRD detectors; thank Shandong Aerospace Electronic Technology Research Institute for assistance in engineering, integrated test and experiment, etc., of detectors; and thank Northwest Rare Metal Materials Research Institute for assistance in supplying beryllium window. This research was supported by the National Natural Science Foundation of China, Grant No. 11775251, the strategic leading science and technology program of Chinese Academy of Sciences (Grant No. XDA 15360100, XDA 15360102).

## References

1. X.B. Han, K.K. Zhang, J. Huang et al., GECAM satellite system design and technological characteristic. Sci. China-Phys. Mech. Astron. **50**, 129507 (2020). https://doi.org/10.1360/SSPMA-2020-0120

2. T.P. Li, S.L. Xiong, S.N. Zhang et al., Insight-HXMT observations of the first binary neutron star merger GW170817. Sci. China-Phys. Mech. Astron. **61**, 031011 (2018). http://arxiv.org/abs/1710.06065

3. E. Bissaldi, A. von Kienlin, G. Lichti et al., Ground-based calibration and characterization of the Fermi gamma-ray burst monitor detectors. Exp. Astron. **24**, 47–88 (2009)

4. S.D. Barthelmy, L.M. Barbier, J.R. Cummings et al., The Burst Alert Telescope (BAT) on the SWIFT Midex Mission. Space Sci. Rev. **120**, 143–164 (2005)

5. A. von Kienlin, N. Arend, G.G. Lichti et al., A GRB detection system using the BGO-Shield of the INTEGRAL-Sectrometer SPI, in *Gamma-Ray Bursts in Afterglow*. ed. by E. Costa, F. Frontera, J. Hjorth (Springer-Verlag, Berlinm, 2001), pp. 427–430

6. R.L. Aptekar, D.D. Frederiks, S.V. Golenetskii et al., Konus-W gamma-ray burst experiment for the GGS Wind spacecraft. Space Sci. Rev. **71**, 265–272 (1995)

7. K. Yamaoka, A. Yoshida, T. Sakamoto, et al. The CALET Gamma-ray Burst Monitor (CGBM). Article 41 in eConf C1304143, (2013)

8. X.Q. Li, X.Y. Wen, Z.H. An et al., The GECAM and its payload. Sci. Sin-Phys. Mech. Astron. **50**, 129508 (2020). https://doi.org/10.1360/SSPMA-2019-0417

9. D.L. Zhang, X.Q. Li, S.L. Xiong et al., Energy response of GECAM gamma-ray detector based on LaBr$_3$: Ce and SiPM array. Nuclear Instrum. Methods A **921**, 8–13 (2019)

10. P. Lv, S.L. Xiong, X.L. Sun et al., A low-energy sensitive compact gamma-ray detector based on LaBr$_3$ and SiPM for GECAM. JINST. **13**, P08014 (2018)

11. R.-Y. Zhu, Precision crystal calorimeters in high energy physics: past, present and future, in *AIP Conference Proceedings* vol. 867, p. 61, (2006), https://doi.org/10.1063/1.2396939

12. A. Papa, P. Schwendimann, Development of new large calorimeter prototypes based on LaBr$_3$(Ce) and LYSO crystals coupled to silicon photomultipliers: A direct comparison. Nuclear Instrum Methods A. **958**, 162999 (2020). https://doi.org/10.1016/j.nima.2019.162999

13. https://www.crystals.saint-gobain.com.

14. F.G.A. Quarati, I.V. Khodyuk, C.W.E. van Eijk et al., Study of 138La radioactive decays using LaBr$_3$ scintillators. NIMA **683**, 46–52 (2012). https://doi.org/10.1016/j.nima.2012.04.066

15. D.Y. Guo, W.X. Peng, Y. Zhu et al., Energy response and in-flight background simulation for GECAM. Sci. Sin-Phys. Mech. Astron. **50**, 129509 (2020). https://doi.org/10.1360/SSPMA-2020-0015

16. R. Klanner, Characterisation of SiPMs. Nuclear Instrum. Methods A **926**, 35–56 (2019). https://doi.org/10.1016/j.nima.2018.11.083

17. https://www.onsemi.com.